# Integrating Quantum Algorithms with Gravitational-Wave Metrology for Enhanced Signal Detection

Vaidik A Sharma
Birla Institute of Technology and Science Pilani, India

**Abstract:-** This study explores the integration of quantum algorithms, specifically Grover's algorithm, with quantum metrology to enhance the efficiency and sensitivity of gravitational-wave detection. By combining quantum matched filtering with precise parameter estimation techniques, the research aims to optimize sensor networks for the identification of gravitational waves. This integrated approach leverages the strengths of quantum superposition and entanglement to improve signal detection, reduce noise, and strategically place sensors. The findings demonstrate significant improvements in the sensitivity and accuracy of gravitational wave measurements, highlighting the potential of quantum technologies to revolutionize observational astronomy and enhance our understanding of the universe.

## I. INTRODUCTION

Exploring the transformative potential of quantum technologies has opened promising avenues in quantum sensing and communication. Quantum computers, although still developing, offer solutions to complex problems beyond classical computers' reach. Alongside quantum computing, Noisy Intermediate-Scale Quantum (NISQ) devices have emerged as significant frontiers. Despite inherent errors, NISQ computers exhibit unique capabilities in various applications like optimization and cryptography. In quantum sensing, they show promise in enhancing detection precision, including potential roles in gravitational wave detection. Gravitational waves, originating from massive cosmic events, present challenges for study. While traditional detectors like LIGO and Virgo are effective, they're costly and have limitations, prompting exploration of more efficient methods, like integrating quantum sensors into networks. Research investigates integrating quantum sensing and communication in NISQ sensor networks for gravitational wave detection. Advanced quantum algorithms such as Quantum Algorithm for Gravitational-Wave Matched Filtering aim to leverage quantum properties for improved detection. Quantum metrology parameter estimation holds promise for navigation, timekeeping, and environmental monitoring. The study addresses practical implications and challenges of deploying quantum-enhanced sensors. The integration of quantum technologies aims to enhance sensitivity and create scalable solutions for understanding mysterious phenomena in the universe.

## II. BACKGROUND

The combination of Quantum algorithm for gravitationalwave matched filtering and quantum metrology presents a comprehensive solution for selecting the most effective quantum sensor network for detecting gravitational waves. By integrating these two techniques, we can improve the network's performance in terms of efficiency and sensitivity, ultimately advancing our understanding of elusive cosmic phenomena.

- Swift Signal Detection: Utilizing Grover's algorithm, the Quantum Algorithm for Gravitational-Wave Matched Filtering accelerates signal detection in noisy data, significantly reducing identification time.
- Precise Parameter Estimation: Quantum metrology complements matched filtering by providing accurate measurements of crucial parameters like frequency, amplitude, and phase associated with gravitational wave signals. This precision enhances the overall accuracy of the detection process.
- Heightened Sensitivity: The synergy between quantum metrology and the quantum algorithm amplifies sensor sensitivity, enabling the detection of even faint gravitational wave signals that may elude classical sensors.
- Noise Reduction: Quantum metrology contributes to noise reduction within the network, ensuring high-quality data collection necessary for accurate template matching in the filtering process.
- Strategic Sensor Placement: Quantum metrology assists in determining optimal sensor placement by estimating parameters related to the gravitational wave source, such as its location and propagation direction, thus optimizing signal detection efficiency.
- Scalability: This integrated approach facilitates the efficient scalability of quantum sensor networks to accommodate diverse research scenarios and evolving scientific requirements.

The integration of advanced quantum techniques holds immense potential in revolutionizing gravitational wave detection. By combining innovative quantum algorithms and precise parameter estimation, we can tailor sensor networks for improved performance, accelerating signal detection, refining measurements, and minimizing noise. Ultimately, this advancement brings us closer to unlocking the mysteries of the universe with greater accuracy and efficiency.

## III. QUANTUM SIMULATION AND METROLOGY TECHNIQUE

➢ *Hamiltonian Simulation and Time Evolution*

At the heart of quantum simulation lies the Hamiltonian operator $H$, which encapsulates the total energy of a quantum system. The time evolution of a quantum state $|\psi(t)\rangle$ governed by the Hamiltonian $H$ is described by the Schrödinger equation:

$$i\hbar \frac{d}{dt}|\psi(t)\rangle = H|\psi(t)\rangle$$

This equation highlights the dynamics of quantum states as they evolve over time under the influence of $H$. Quantum computers excel at simulating this time evolution, enabling the study of quantum systems that are challenging or impossible to simulate using classical methods [1]. The ability to simulate quantum systems holds promise for tasks such as understanding chemical reactions, optimizing materials, simulating condensed matter systems and gravitational wave astronomy [11].

➢ *The Time Block Method: Mitigating Simulation Errors*

In quantum simulation, the time evolution is often approximated through discrete time steps using the Trotter-Suzuki decomposition:

$$e^{-iHt} \approx \left(e^{-iH\delta t}\right)^{T/\delta t}$$

Where $T$ is the total evolution time and $\delta t$ is the time step. The Trotter-Suzuki approximation breaks down the continuous evolution into a sequence of smaller steps, simplifying the simulation process. However, long simulations can accumulate errors from each time step.[1]

The upper bound of the error introduced by the time block method can be estimated using the Lie-Trotter formula, which provides an expression for the difference between the exact time evolution and the approximation:

$$\left\| e^{-iHt} - \left(e^{-iH\delta t}\right)^{N_{\text{blocks}}} \right\| \leq \frac{\| H \|^2 T^3}{3\hbar} \left(\frac{T}{N_{\text{blocks}} \hbar}\right)^{2k-1}$$

Where $\| H \|$ is the operator norm of $H$ and $k$ is the order of the Lie-Trotter formula. This derivation allows us to quantify the accuracy of the time block method and optimize the choice of time step, number of blocks, and order of the formula for a given simulation.

➢ *Variational Parameter Estimation and Quantum Metrology*

Variational parameter estimation is a powerful technique in quantum computation,[2] particularly in the context of metrology. Quantum metrology leverages the principles of quantum superposition and entanglement to achieve measurements with higher precision than classical methods allow. Variational circuits, parameterized by angles $\theta$, can be used to prepare quantum states and perform measurements. The goal is to find the optimal parameters that minimize a loss function $L(\theta)$ and produce the desired quantum state.[5],[6]

Mathematically, the variational parameter estimation process can be formulated as an optimization problem:

$$\theta^* = \underset{\theta}{\operatorname{argmin}} L(\theta).$$

Quantum metrology techniques use the optimized parameters to enhance the accuracy of parameter estimation.[2] The Fisher information $(F)$ quantifies the sensitivity of the quantum state to variations in the parameter $\theta$. In bra-ket notation, it is given by:

$$F(\theta) = \sum_k \frac{\left|\left\langle \psi_k \left| \frac{d}{d\theta}\psi(\theta) \right\rangle\right|^2}{p_k}$$

Where $p_k$ is the probability of outcome $k$ and $\frac{d}{d\theta}\psi(\theta)$ is the derivative of the quantum state with respect to $\theta$. The Fisher information sets a fundamental limit on how precisely a parameter can be estimated.[3]

The Fisher information is intimately connected to the expectation of the score operator ( $S$ ), which is defined as the derivative of the logarithm of the likelihood function:

$$S(\theta) = \frac{1}{\sqrt{p(\theta)}} \frac{d}{d\theta} \sqrt{p(\theta)}$$

The Fisher information can be expressed as the variance of the score operator:

$$F(\theta) = \operatorname{Var}(S(\theta))$$

This relationship highlights the role of the Fisher information in quantifying the information content of the measurements with respect to the parameter $\theta$.

The Cramer-Rao bound provides a mathematical relation between the Fisher information and the achievable precision of parameter estimation. For an unbiased estimator $\hat{\theta}$, the CramerRao bound states:

$$\operatorname{Var}(\hat{\theta}) \geq \frac{1}{NF(\theta)}$$

Where $\text{Var}(\hat{\theta})$ is the variance of the estimator and $N$ is the number of measurements.

Quantum metrology techniques aim to approach the CramerRao bound by optimizing measurement strategies and exploiting quantum entanglement to enhance the Fisher information. This enables quantum systems to achieve measurements with unprecedented precision, surpassing classical limits.

➢ *Four-Step Procedure for Information Extraction*

The four-step procedure for information extraction in quantum computation involves a systematic approach to preparing quantum states, evolving parameterized quantum states, measuring outputs, and estimating parameters based on multiple measurements. This procedure is integral to variational parameter estimation and quantum metrology, enabling the enhancement of predictive accuracy and precision in quantum computations.

- Preparation of Input States: The first step of the procedure involves the preparation of input quantum states. These states serve as the initial conditions for the quantum computation. Parameterized quantum circuits are used to generate these states, where the parameters $\theta$ determine the quantum state's characteristics (Fig.1. and Fig.2.). Variational techniques are applied to optimize these parameters, ensuring that the prepared states are tailored to the specific problem at hand.
- Evolution of Parameterized Quantum States: Once the input states are prepared, the next step is to evolve them over time using the Hamiltonian operator $H$. This time evolution is achieved through the application of quantum gates that implement the unitary operator $U(t) = e^{-iHt/\hbar}$. The parameterized nature of the quantum circuit allows for flexibility in controlling the evolution dynamics. The optimization of parameters using variational methods ensures that the quantum evolution approximates the desired transformation accurately.
- Measurement of Outputs: Following the evolution of quantum states, measurements are performed to extract relevant information. Observable quantities, represented by Hermitian operators, are measured to obtain measurement outcomes. These outcomes provide insights into the quantum system's behavior and dynamics. Quantum measurements introduce inherent randomness due to the probabilistic nature of quantum states, requiring multiple repetitions to gather sufficient statistical data.[4]
- Estimation of Parameters: The final step of the procedure involves the estimation of parameters based on the measurement outcomes. Estimators are used to infer the optimal parameter values that best align with the obtained measurements. Variational optimization techniques, such as gradient descent, are commonly employed to minimize the difference between the observed outcomes and the predicted outcomes from the parameterized quantum circuit. This iterative process refines the parameter estimates, leading to improved accuracy and predictive power.
- The four-step procedure for information extraction serves as a fundamental framework in quantum computation, encompassing the key stages of preparing states, evolving quantum dynamics, measuring observables, and optimizing parameters. This systematic approach underlies the advancements in variational quantum algorithms and quantum metrology, driving the development of accurate and precise quantum predictions.

➢ *Ramsey Interferometer Quantum Circuit in Experimental Setup*

In experimental quantum metrology, the Ramsey interferometer quantum circuit plays a pivotal role. The Ramsey interferometer is a fundamental quantum device used to measure frequency and phase shifts with exceptional precision [5]. It consists of two sequential applications of a $\pi/2$ pulse separated by a time delay $T$ and followed by a final $\pi/2$ pulse. This configuration effectively splits the quantum state into two branches, allowing interference between the branches after the second pulse. By varying the time delay $T$, the Ramsey interferometer becomes sensitive to small changes in frequency or phase.[5] Mathematically, the Ramsey interferometer can be represented as a sequence of unitary operators. Let $U_{\pi/2}$ be the unitary operator corresponding to a $\pi/2$ pulse and $U_T$ be the unitary operator corresponding to the time delay $T$. The Ramsey interferometer circuit can be described as:

$$\text{Ramsey Circuit} = U_{\pi/2} \cdot U_T \cdot U_{\pi/2}$$

The unitary operators $U_{\pi/2}$ and $U_T$ can be represented in matrix form, where $H$ is the Hadamard gate and $e^{-iHt}$ is the time evolution operator with the Hamiltonian $H$ over time :

$$U_{\pi/2} = H \cdot e^{-i\frac{\pi}{4}}$$
$$U_T = e^{-iHT}$$

The interferometer's output state after the second $\pi/2$ pulse can be obtained by applying the Ramsey circuit to the initial quantum state $|\psi\rangle$ :

$$|\psi_{\text{out}}\rangle = \text{Ramsey Circuit} \cdot |\psi\rangle$$

The resulting state $|\psi_{\text{out}}\rangle$ exhibits oscillatory behavior as a function of the time delay $T$, allowing for the measurement of phase shifts with high precision.

Integrating the Ramsey interferometer quantum circuit into the four-step procedure enhances the capabilities of quantum metrology. The interferometer's sensitivity to phase shifts makes it a valuable tool for applications such as atomic clocks, quantum sensors, and precision measurements in fundamental physics.

The four-step procedure, combined with the versatile Ramsey interferometer, exemplifies the power of variational quantum algorithms and quantum metrology. This approach facilitates accurate predictions and measurements, with the potential to revolutionize fields reliant on high-precision data.

## IV. GRAVITATIONAL WAVE MATCHED FILTERING

We'll utilize the PyCBC library, a tool designed for analyzing gravitational-wave data, identifying astrophysical sources from compact binary mergers, and examining their characteristics. These tools mirror those employed by the LIGO and Virgo collaborations for detecting gravitational waves within their data.[10]

We generate the waveform of a gravitational-wave merger and matched filtering, which is optimal in the case of Gaussian noise and a known signal model. In reality our noise is not entirely Gaussian, and in practice we use a variety of techniques to separate signals from noise in addition to the use of the matched filter. Here we generate the gravitational waveform using one of the available waveform approximants. These can be generated as a time series using get_td_waveform. The essential factors include the masses of the binary system (measured in solar masses), the sampling interval (in seconds), the initial frequency of gravitational waves (in Hz), and the choice of approximation method. Numerous approximation methods are accessible, each accounting for various physical phenomena.

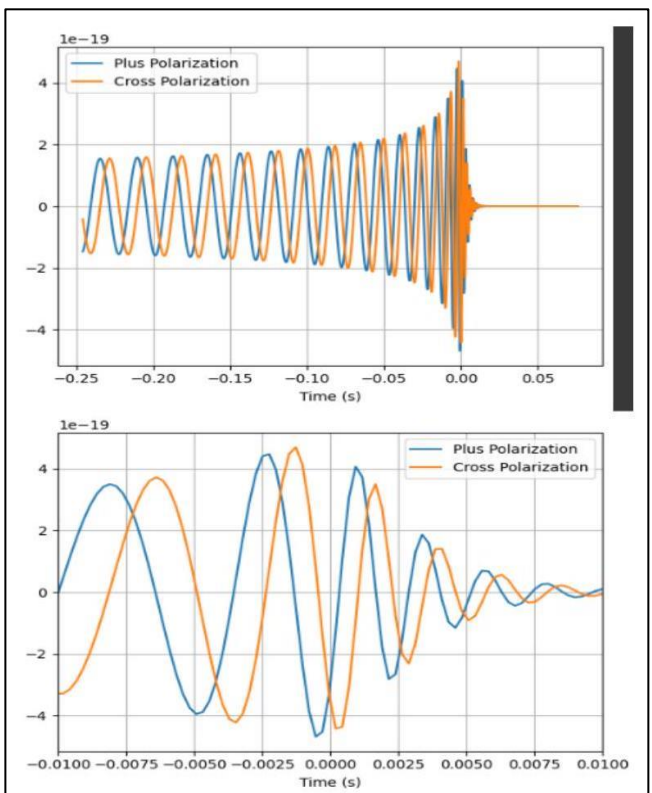

Fig 1 Waveform Generated

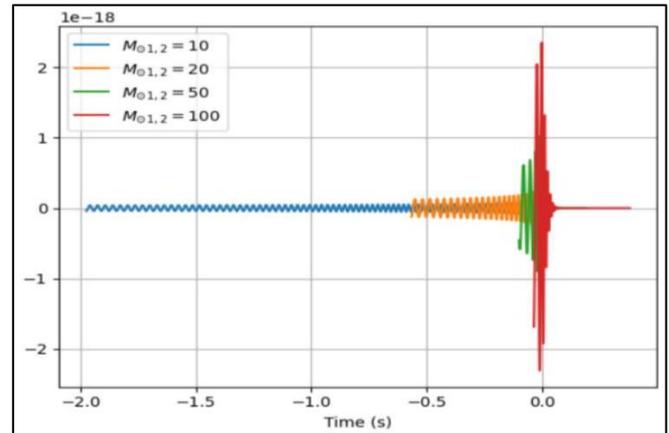

Fig 2 Waveform Corresponding Different Masses

Here, we've chosen to use the 'SEOBNRv4_opt' model. Numerous alternatives are accessible, each employing distinct approaches and encompassing varied physical phenomena. This particular model simulates the gravitational waveform produced by the merging of black holes, allowing for the spin of each black hole to align with the orbit. The specified parameters are: mass $1 = 20$, mass $2 = 20$, delta_$t = 1.0/4096$, f_lower=40.. (**fig**. 1)

It can be compared that the length of the waveform increases for lower mass binary mergers.(Fig. 2)

The distance of the waveform follows a simple linear relation between distance (in megaparsec or Mpc) and apmplitude when red-shift is not taken in consideration.(Fig. 3)

Then we reduced the dynamic range of the data and supress low freqeuncy behavior which can introduce numerical artefacts. We also downsampled the data to 2048 Hz as high frequency content is not important.(Fig. 4)

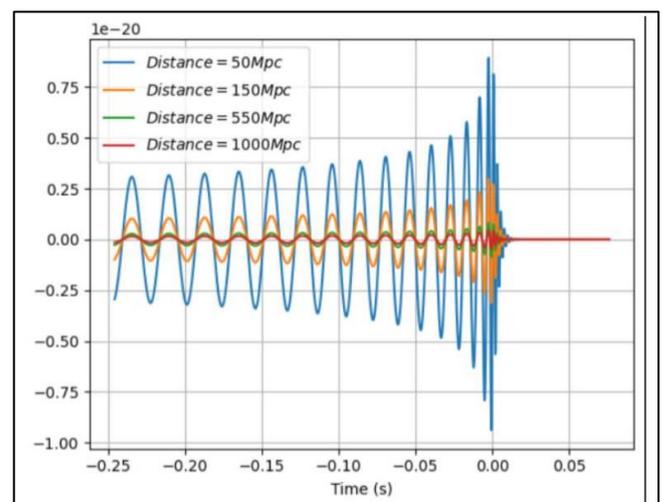

Fig 3 Waveform Corresponding Different Distances

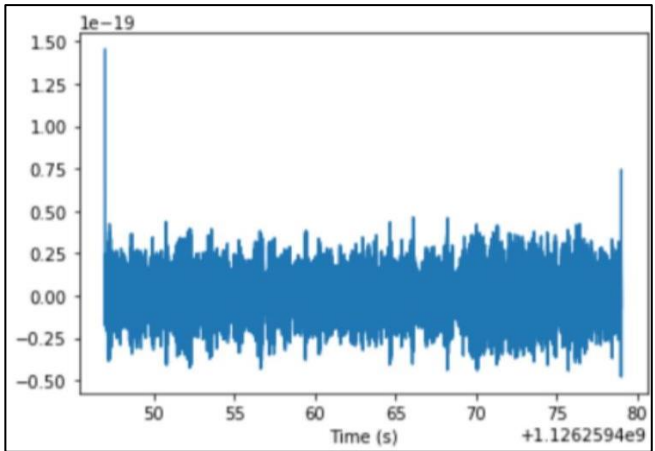

Fig 4 H1 Recorded GW150914 Merger

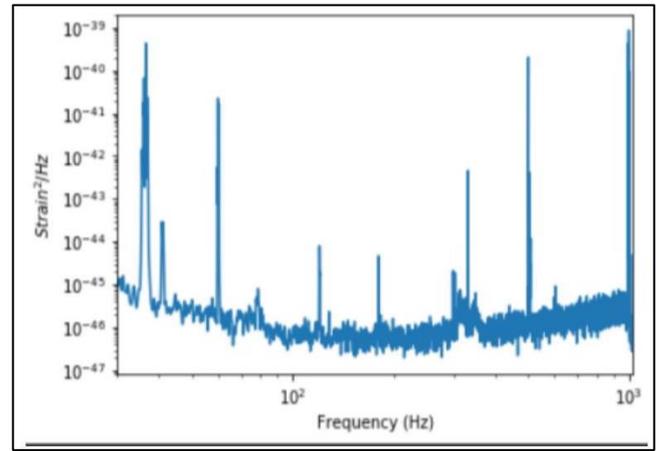

Fig 6 Spectral Power Density

Observing a notable increase in the data at its edges, we attribute this to the impact of the highpass and resampling processes employed in filtering. As the filter is applied to the edges, it loops back to the start of the data, resulting in a spike. This phenomenon occurs because the data lacks cyclic continuity, causing the filter to resonate for a duration equivalent to its length. While any visible transients may not be apparent, it's imperative to avoid filters operating on non-causally connected times. To mitigate this issue, we opt to trim the data ends adequately to prevent wrapping around the input. This requirement will be consistently enforced across all filtering stages. (Fig. 5)

Effective matched filtering involves adjusting the weighting of frequency components in both the potential signal and the data according to the noise amplitude. This process can be likened to filtering the data using a time series version of the reciprocal of the Power Spectral Density (PSD). To ensure control over the extent of this filtering, we apply a windowing technique to the time domain equivalent of the PSD, limiting its length. While this approach may result in some loss of information regarding line behavior within the detector, the impact is minimal due to the broad frequency range covered by our signals and the narrowness of the lines.(Fig. 6)

Conceptually, matched filtering involves laying the potential signal over your data and integrating (after weighting frequencies correctly). If there is a signal in the data that aligns with the 'template', you will get a large value when integrated over. (Fig. 7)

Then SNR-signal to noise time series is calculated. (Fig. 8) In our previous analysis, we identified a peak in the signalto-noise ratio (SNR) concerning a suggested merger of binary black holes. Utilizing this SNR peak, we aim to align our proposal with the actual data and also to eliminate our proposal's influence from the data. To ensure a fair comparison between the data and the signal, and to focus on the relevant frequency range, we intend to standardize both the template and the data through a process known as whitening. Subsequently, we will filter both the data and the template within the frequency range of 30 to 300 Hz. By doing so, any signal present in the data will undergo the same transformation as the template, facilitating a more equitable comparison. (Fig. 9)

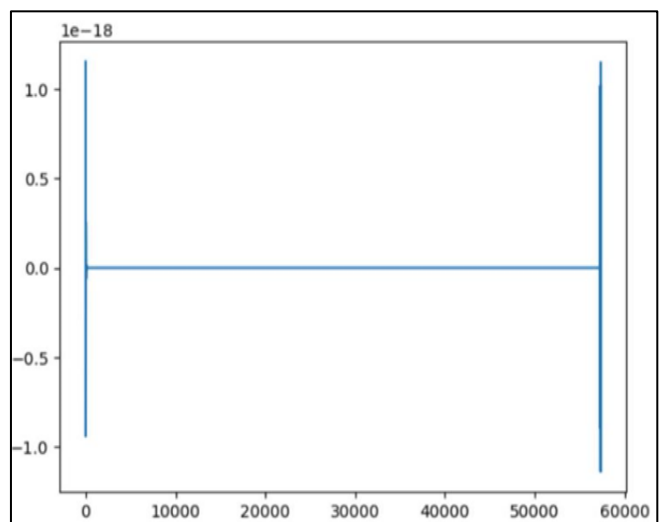

Fig 7 Signal Model

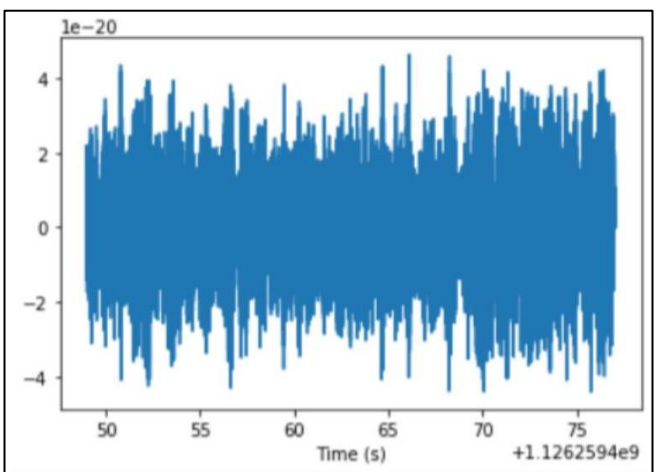

Fig 5 Filter Wraparound

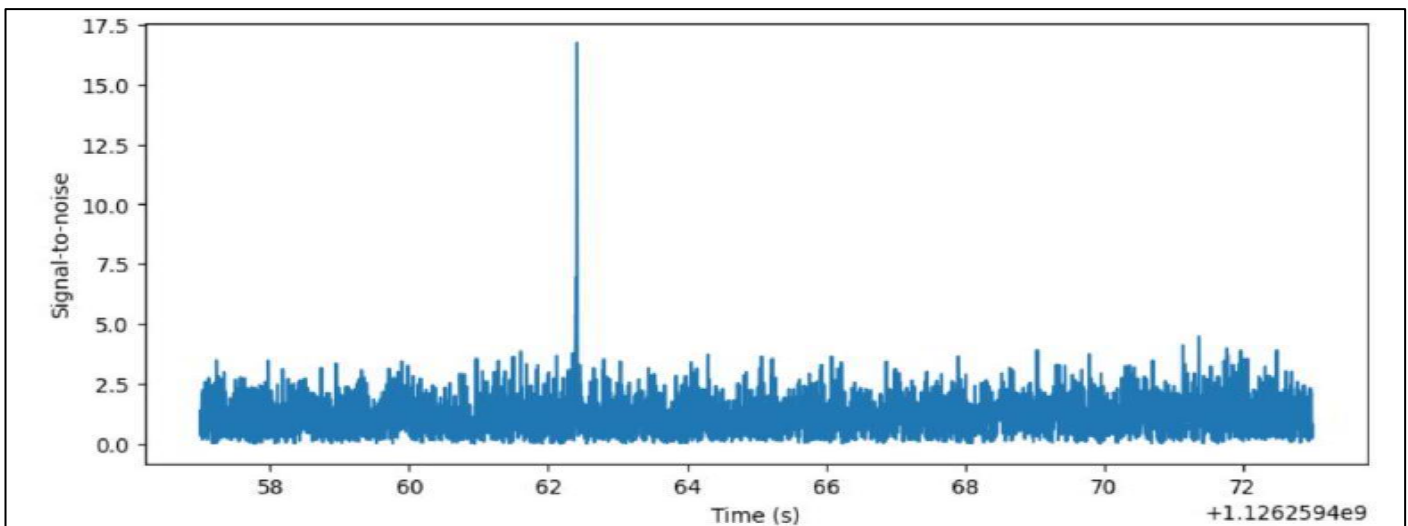

Fig 8 Signal to Noise Ratio

After alginment we subtract the template from original H1 data and results are obtained. (Fig. 10)

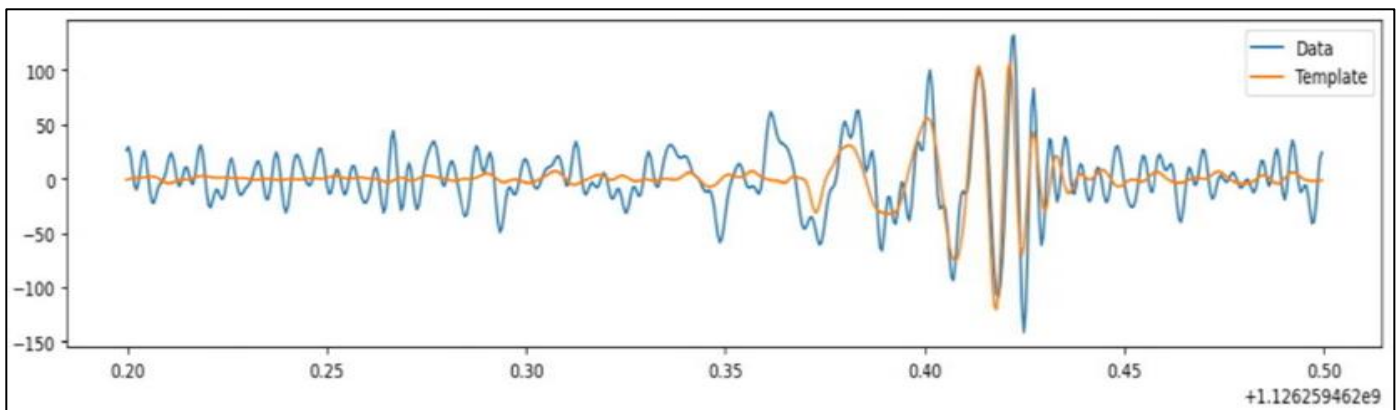

Fig 9 Data v/s Template

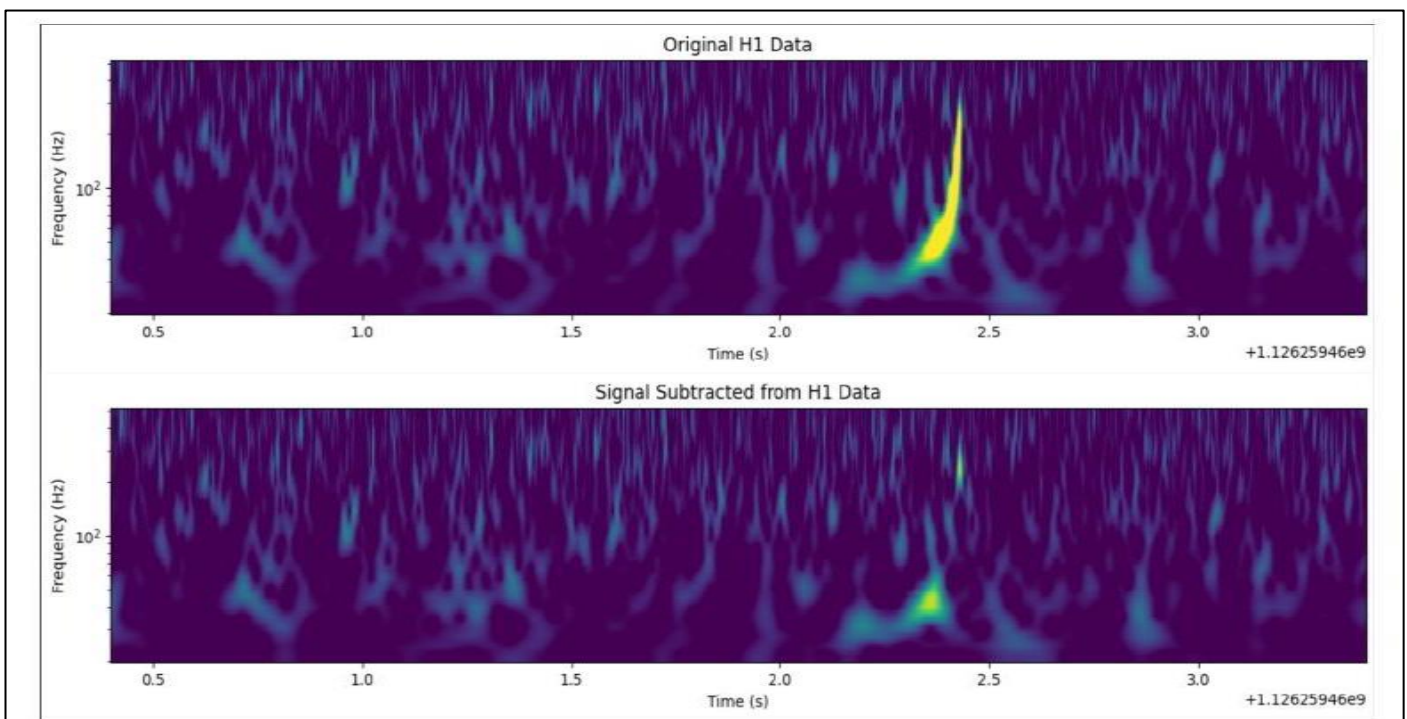

Fig 10 Original and Subtracted Signal

## V. QUANTUM GRAVITATIONAL WAVE MATCHED FILTERING

As previously explained, matched filtering involves comparing time series data with templates to identify matches above a specified threshold. Gravitational wave (GW) data templates, which adhere to general relativity principles, are computed on the fly rather than being preloaded from a database, as is often impractical due to their sheer number. In a quantum context, this eliminates the need for a massive data transfer into a quantum random-access memory (qRAM). The process of constructing an oracle to determine template matches is part of classical data analysis and can be explicitly integrated without diminishing the quantum approach's speedup potential. However, it's crucial to note that the computational cost of an oracle call, typically involving signal-to-noise ratio (SNR) calculations, remains significant. Grover's algorithm doesn't expedite this step. Still, quantum counting improves the computational cost's dependence on the number of templates, making previously unmanageable searches feasible. Importantly, quantum counting [9] enables the detection of extremely faint signals that classical methods can't discern, enhancing the effectiveness of matched filtering in gravitational wave data analysis.[11],[12],[13]

➢ *Quantum Parameter Estimation*

Estimators are used to infer the optimal parameter values that best align with the obtained measurements. Variational optimization techniques, such as gradient descent, are commonly employed to minimize the difference between the observed outcomes and the predicted outcomes from the parameterized quantum circuit. This iterative process refines the parameter estimates, leading to improved accuracy and predictive power.[4]

The need for careful LR tuning to ensure accurate predictions and robust training in parameterized quantum circuits is observed. Additionally, the quantum advantage demonstrated by the parameterized quantum circuits, showcases the potential of quantum computation techniques in revolutionizing predictive modeling and achieving unprecedented levels of accuracy and precision.

The 3-qubit quantum network consistently outperformed the 2-qubit network with stable convergence, highlighting the benefits of increased qubit complexity in capturing intricate data patterns. This study emphasizes the significance of tailored hyperparameters and model architectures. While quantum networks offer intriguing potential, meticulous tuning is crucial.

➢ *Oracle Construction*

To employ quantum counting in these applications, we initiate the process with the creation of an oracle, which performs matched filtering while adhering to a predefined threshold. The initial step involves outlining the pseudocode for the construction of Grover's gate.[14]

In order to understand the framework, we begin with some essential preliminaries: N represents the number of templates, while M signifies the number of data points in the time series. The data and templates are digitally encoded as classical bits in the computational basis. Converting a classical, irreversible logic circuit into a reversible one is feasible, and this can be readily implemented on a quantum computer by substituting classical reversible gates with their quantum equivalents. This procedure demands some scratch space to facilitate reversible.

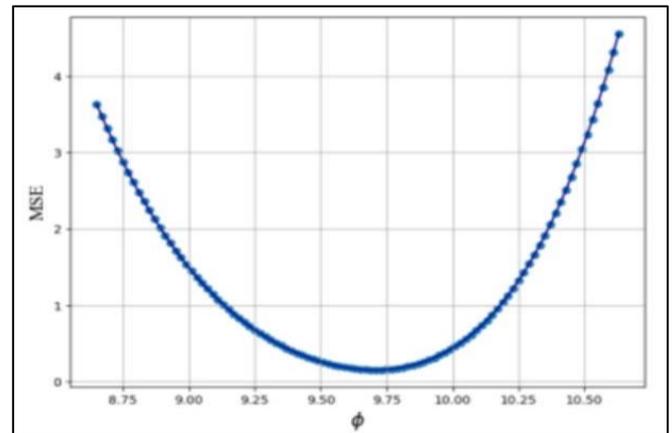

Fig 11 2 Qubits PQC Result-1

MSE vs $\phi$ plot for 2 Qubits Circuit, Learning Rate = 0.01

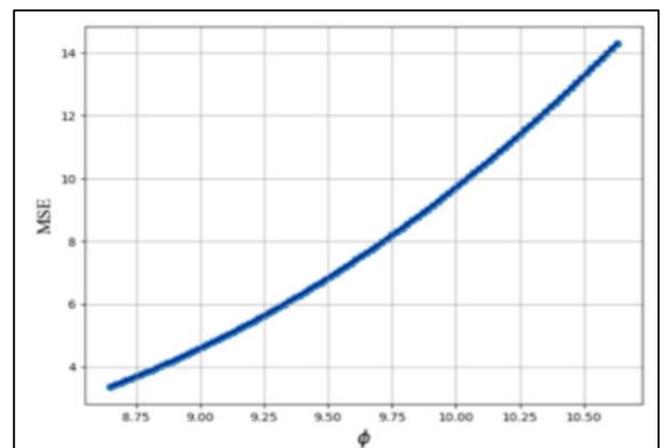

Fig 12 2 Qubits PQC Result-2

MSE vs $\phi$ plot for 2 Qubits Circuit, Learning Rate = 0.04

➢ *Calculations.*

The implementation involves the use of four registers: a data register (sized linearly with M), an index register (requiring $\log_2 N$ qubits), a template register (sized linearly with M for intermediate calculations), and an SNR register (with a constant size, $O(1)$).

The core of Grover's algorithm is the search over an index within a database, which necessitates the construction of an oracle. The oracle calculates the template from an index i, computes the SNR, and checks the result against a predefined threshold. The number of gates required to compute a template waveform from its parameters is denoted as k1, and it scales linearly with M. The calculation of the SNR between a template and the data requires k2 gates, with a time complexity of $O(M \log M)$. Finally, verifying whether the result surpasses a given threshold $\rho$ thr, necessitating $O(1)$ gates, is denoted as k3. The overall computational complexity of the classical algorithm for computing matches against all templates is $O(NM \log M)$.

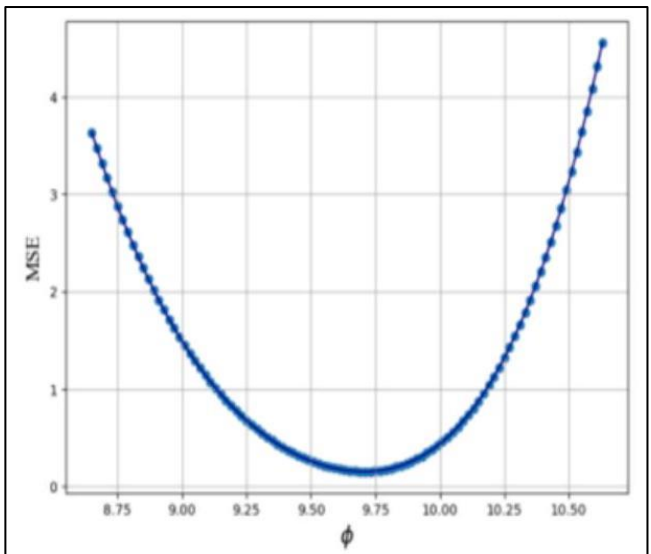

Fig. 13 3 Qubits PQC Result-1

MSE vs $\phi$ plot for 2 Qubits Circuit, Learning Rate = 0.01

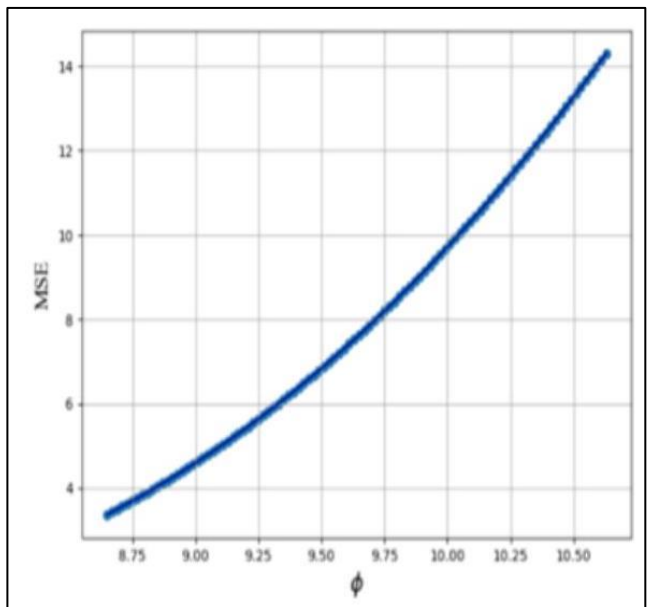

Fig 14 3 Qubits PQC Result-2

MSE vs $\phi$ plot for 2 Qubits Circuit, Learning Rate = 0.04

- *Step 0 (Initialization) [Cost: $O(M + \log N)$ ]:*
  This step initializes the initial state, composed of four registers, and requires loading data, which takes linear time in M, and initializing the index register to an equal superposition, which needs $O(\log N)$ gates.

- *Step 1 (Creating Templates) [Cost: $O(M)$ ]:*
  Calculating templates from the index is performed in superposition over all index values, with a cost of k1 approximately $O(M)$ gates. Step 2 (Comparison with the Data) [Cost: $O(M \log M)$ ): Calculating the SNR between the template and the data costs k2 approximately $O(M \log M)$. It is then compared to a threshold to determine f(i), which denotes whether a template is a match, with a cost of k3 approximately $O(1)$.

- *Step 3 (Disentangling Registers) [Cost: $O(M \log M)$ ]:*
  This step involves erasing intermediate calculations to disentangle the index register from the other registers. The erasure process has a cost of $k1 + k2$, which is roughly $O(M \log M)$.

- *Step 4 (Applying the Diffusion Operator) [Cost: $O(\log N)$] :*
  This step, unique to the quantum algorithm, requires $O(\log N)$ quantum gates.

In total, the cost for a single oracle call is $O(M \log M + \log N)$. The integration of quantum counting in GW matched filtering introduces efficiencies in match determination and template retrieval. The construction of a quantum oracle for this purpose enhances the computational process, ultimately improving the efficiency of the algorithm.

➢ *Signal Detection*
The primary focus is on signal detection, involving four conditional probabilities.

- True Negative $(P(r^* = 0 \mid r = 0))$ :
  The probability of correctly determining that no template exceeds the SNR threshold when there is no such template.

- False Negative $(P(r^* = 0 \mid r > 0))$ :
  The probability of incorrectly identifying no match when, in reality, no template exceeds the SNR threshold.

- True Positive $(P(r^* > 0 \mid r > 0))$ :
  The probability of correctly identifying templates exceeding the SNR threshold when such templates exist.

- False Alarm $(P(r^* > 0 \mid r = 0))$ :
  The probability of incorrectly identifying templates exceeding the SNR threshold when no such templates exist.

These probabilities differ from traditional definitions, as they account for classification errors caused by the probabilistic nature of quantum algorithms. These classification errors are our main focus in this study.

➢ *Quantum Counting Procedure:*

- Step 1: Initialization [Cost: $O(M + \log N)$] - Setting the stage with data loading and index register initialization.
- Step 2: Creating Counting Register [Cost: $O(1/2 \log N)$] – Applying Hadamard gates to qubits, incurring a cost of p. Step 3: Controlled Grover's Gate [Cost: $O((M \log M + \log N)/\sqrt{N})$] - Determining the maximum iterations of Grover's gate required $(2p - 1)$.
- Step 4: Inverse Quantum Fourier Transform [ Cost: $O((\log N)^2)$].
- Step 5: Measurement [Cost: $O(1/2 \log N)$] - Calculating b, the measurement outcome.

The overall cost of the algorithm is $O(\sqrt{N}(M \log M + \log N))$.

The choice of 'p' depends on the desired accuracy and is related to the probability of false negatives $(\delta n)$. A specific value of 'p' is recommended to minimize false negatives while controlling computational cost.

The algorithm's cost is significantly lower than classical approaches, making it efficient for signal detection tasks involving a large number of templates (e.g., in gravitational wave research). The computational cost difference is particularly prominent in cases with numerous templates.[7],[11]

➢ *Matched Templates Retrieval*

The process of retrieving matching templates relies on Grover's algorithm from Algorithm 1 and utilizes the estimated number of matching templates ('r*') from Algorithm 2. While there are multiple approaches to retrieve matching templates when the number of matches is unknown, this algorithm presumes that the signal detection algorithm is executed first, and its estimate of '$r$''' is naturally used for subsequent retrieval attempts.

- *Algorithm 3 - Template Retrieval:*

✓ *Calculating the Number of Repetitions:*
The initial step involves calculating the number of repetitions required to retrieve the desired template (cost: $O(1)$ ). The value ' $r^*$ ' ob − tained from Algorithm 2 is used to determine the number of repetitions based on a mathematical expression.

✓ *Template Retrieval Complexity:*
The algorithm's complexity is described as $O(\sqrt{N}(M \log M + \log N)\sqrt{N})$, considering Algorithm 2 and the retrieval of one template together.

- *Procedure 1 (Retrieve One Template):*
Repeat Grover's algorithm (Algorithm 1) ' $k^*$ ' times to obtain the desired template index (cost: $O(\sqrt{N}/r * (M \log M + \log N))$ ). The value of ' $k^*$ ' is derived from previous discussions and is proportional to $\sqrt{N}/r^*$.

- *Procedure 2 (Retrieve All Matched Templates):*
If the goal is to retrieve all matched templates, it's not as straightforward as Procedure 1. It's akin to a "coupon collector problem," which requires approximately ( $\log r$) repetitions of Procedure 1. The complexity is similar for both procedures, provided the number of matching templates is significantly smaller than the total number of templates.

- *Probability of Failure:*
The section concludes by discussing the overall probability of failing to retrieve a matched template using this procedure. It's emphasized that if this probability is less than 0.5 , then with a finite number of repetitions, it can be reduced to a negligibly small value, ensuring successful retrieval of a matched template.

➢ *E. Grover's Algorithm and Matched Filtering*
We try to find the frequency of a sine wave signal from amongst a number of known frequencies when the signal start time and amplitudes are known. Time domain data D of length N is taken from the signal and compared to M templates T of sine waves of known frequencies equally spaced out in the frequency space of interest. The number of times Grover's algorithm is applied is dependent on P.

The signal sin wave looks like Fig 15.

First let us make the state:

$$|\psi_{\text{ini}}\rangle = \frac{1}{\sqrt{M}} \sum_{m=0}^{M-1} |m\rangle|0\rangle^{\times N}$$

This state is what the template information will be stored in. Let us also define:

$$|i\rangle = \frac{1}{\sqrt{M}} \sum_{m=0}^{M-1} |m\rangle$$

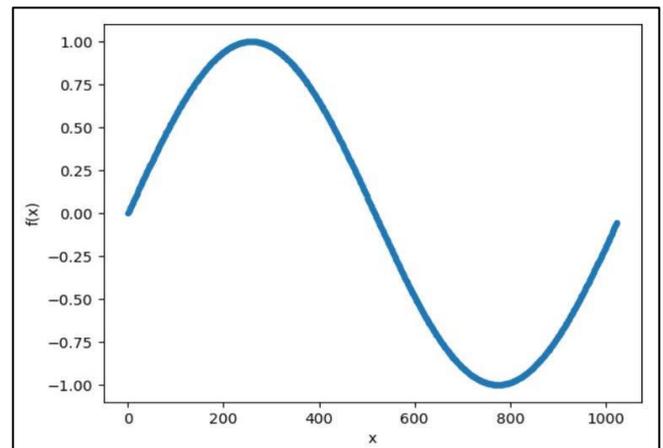

Fig 15 Sin Signal

This state has a basis state that corresponds to the indexes of each template. We perform operation $\hat{k}_1$ on this state to give a state that represent the template waveforms in the frequency domain in $|T\rangle$ :

$$|i\rangle \otimes |T\rangle = \hat{k}_1(|i\rangle \otimes |\psi_{\text{ini}}\rangle)$$

Similarly the data is loaded into a state represented by $|D\rangle$. We get the template plot as shown in Fig. 16.

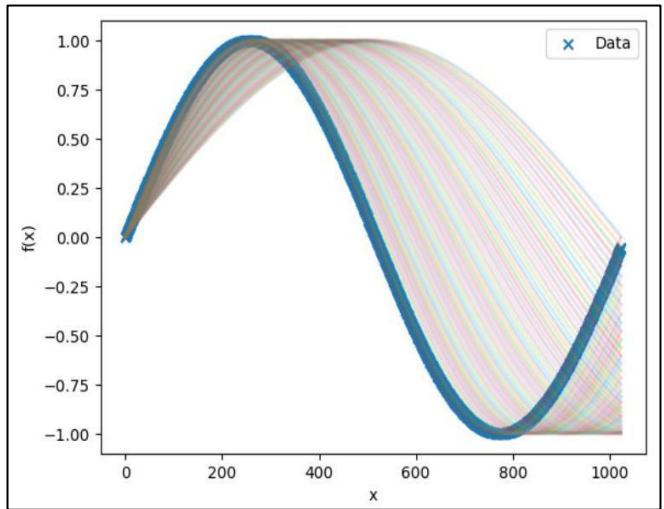

Fig 16 Templates

Grover's algorithm is applied after matched filtering is to all templates. On a quantum computer, this can be done in parallel to give an equal amplitude state $|w\rangle$ of length $M$ but any state that corresponds to the index of a template that meets the criteria of $|T - D| < 10^{-3}$ has a phaseflip of -1. $|w\rangle$ is made by applying $\hat{k}_2$ :

$$|i\rangle \otimes |T\rangle \otimes |D\rangle \otimes |w\rangle = \hat{k}_2(|i\rangle \otimes |T\rangle \otimes |D\rangle \otimes |0\rangle^{\times 2^M}).$$

Then we make the state:

$$|\psi_0\rangle = \frac{1}{\sqrt{PM}} \sum_{p=0}^{P-1} \sum_{m=0}^{M-1} |p\rangle|m\rangle$$

This is equivalent of the state $|s\rangle$ previously made, but $P$ times.

- *The first part of Grover's algorithm is then applied as follows:*

✓ Create operator $\hat{U}_w = I - 2|w\rangle\langle w|$ where $w$ is the matrix position corresponding to the matching templates. This operator has the property:

$$\hat{U}_w|x\rangle = -|x\rangle \text{ if } x = w$$
$$\hat{U}_w|x\rangle = |x\rangle \text{ if } x \neq w$$

✓ Initiate superposition:

$$|s\rangle = \frac{1}{\sqrt{M}} \sum_{x=0}^{M-1} |x\rangle$$

Assuming that every template is equally likely to have the correct template without any more prior knowledge.

✓ Create the Grover diffusion operator $\hat{U}_s = 2|s\rangle\langle s| - I$.
✓ Apply $\hat{U}_w$ then $\hat{U}_s$ to $|s\rangle p$ times to each state in $P(p = \{0,1,\dots P-1\})$. Now we apply Grover's algorithm itterably to this state such that:

$$|\psi_1\rangle = \hat{C}_G|\psi_0\rangle$$

Where $\hat{C}_G|p\rangle \otimes |m\rangle \to |p\rangle \otimes (\hat{G})^p|m\rangle$

and $\hat{G} = \hat{U}_s \hat{U}_w$

We can see what this operation does to $|\psi_0\rangle$ in the plot Fig. 17.

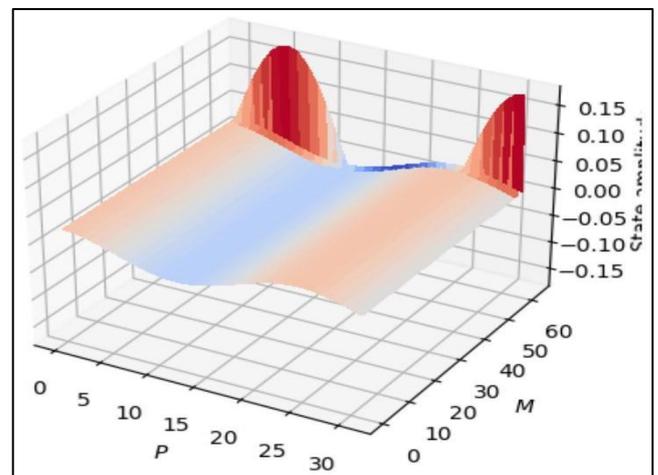

Fig 17 State Probability

We see the amplitudes of states $\psi_1$ numbering $P \times M$. Where the amplitudes of the states around correct solutions are much greater than the incorrect solutions. The incorrect solutions also exhibit a sinusoidal pattern.

It is the frequency $f$ of this sinusoid that we wish to determine, as it is related to the number of correct template matches $k$ by:

$$k = \sin^2 \frac{f\pi}{P}$$

Determining phase/frequency information from amplitudes of states requires a quantum fourier transform. The quantum fourier transform is much the same as it's classical counterpart but is performed on amplitude/phase information stored on the states of qubits. It transfers information stored in amplitudes in quantum states into phase information. There also exists the inverse quantum fourier transform for the reverse opperation. The quantum fourier transform acting on state $|p\rangle$ gives:

$$\text{QFT}: |x\rangle \mapsto \frac{1}{\sqrt{K}} \sum_{k=0}^{K-1} e^{2\pi i \frac{kx}{K}} |k\rangle$$

The inverse quantum fourier transform is applied across the ancillary qubits the recover the phase information from the sinusoidal behaviour in the states shown in the graph above. This requires creating a $QFT^{-1}$ operator of size $P \times P$, which we will call $\hat{F}_P$ as done below:

Applying this across the ancillary states in $|\psi_1\rangle$ :
$$\psi_2 = \hat{F}_P \psi_1$$

This operation changes $|\psi_1\rangle$ to the states in the plot (Fig. 18)

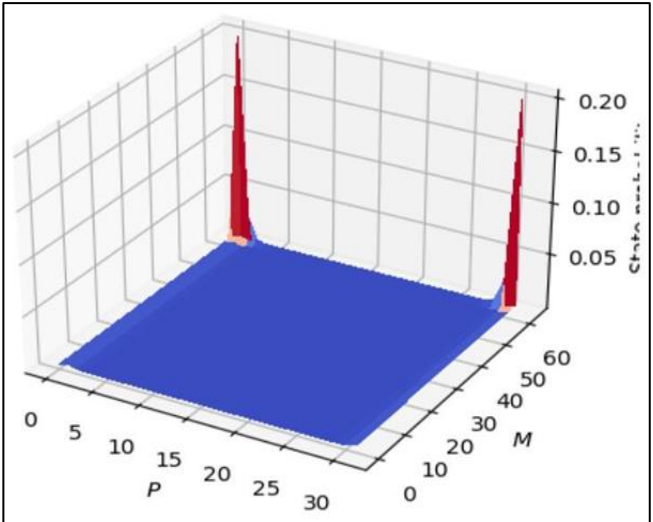

Fig 18 New State Probability Distribution

The probabilities $|\psi_2|^2$ are seen over ancillary and template states. There are two peaks, corresponding to $f$ and $P - f$. Measuring the ancillary state will likely acquire one or the other. These peaks also correspond to the positions of matching templates, and so measuring the template qubits likely recovers a correct position of a template.

From knowing $f$ or $P - f$ we can work out the number of matched templates $k$ with the relation:

$$k = M\sin^2\frac{f\pi}{P}$$

Doing this gives the number of matching templates.

If the number is greater than $0$ , we have matching templates.

We find a corresponding matching template. To do this we need to find out the optimum number of Grover's applications to apply. This can easily be found from knowing the number of matching templates:

$$p_{\text{opt}} \approx \frac{\pi}{4}\sqrt{\frac{M}{k}}$$

Now we can just apply Grover's algorithm $p_{\text{opt}}$ times to $\frac{1}{\sqrt{M}}\sum_{i=0}^{M-1}|m\rangle$ to result in a state with amplitudes maximally amplified corresponding to matching states: We observe how the maximum state probability changes over applications of Grover's iterations. This is plotted (Fig. 19) along with a line indicating the determined optimal number of applications.

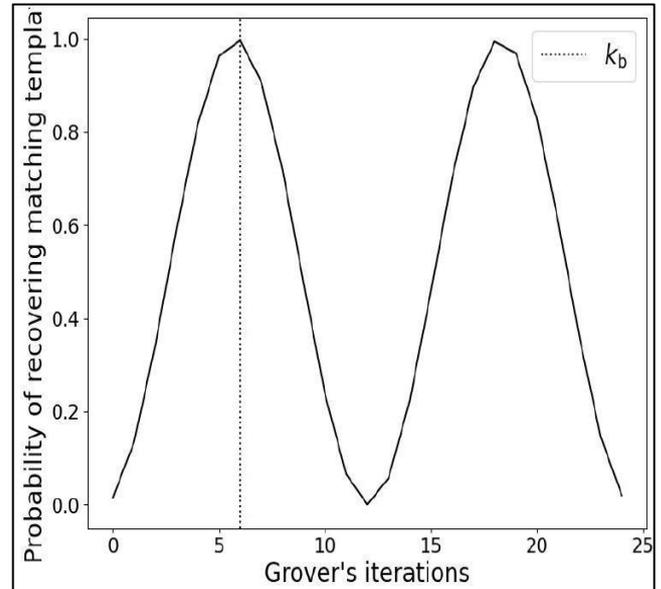

Fig 19 State Probability over Grover Iterations

➢ *F. Qiskit Implementation*
Firstly, the registers are initialized. This is to set the data we are matching against.

In the case of have multiple matching templates, we shall not use 'search_Circuit.h(data[0])'. This will create two data that we are matching against, rather than two matching templates. In this case, both data would run as superpostions and so will the matching result. In essense, it is still one template searching so no untilization of quantum counting.

The Z-gate on the counting qubit is to compensate the general $\phi$ phase introduced by the Diffusion operator function.
Secondly, Grover's gate and Oracle function is created.The oracle works by matching data and template bit by bit in the qubit range within the precision. We use a C-X gate for matching, in which case, the template qubits would be, and only be in state 10⟩ if it is a match. After bit flipping, by applying a C-X gate ctrolled by all the template qubits, the ancilla qubit would be and only be flipped if all templates qubits matched. After the matching, we need to reverse all the previous actions on the templates qubits.

Then, Diffusion operator function is created.

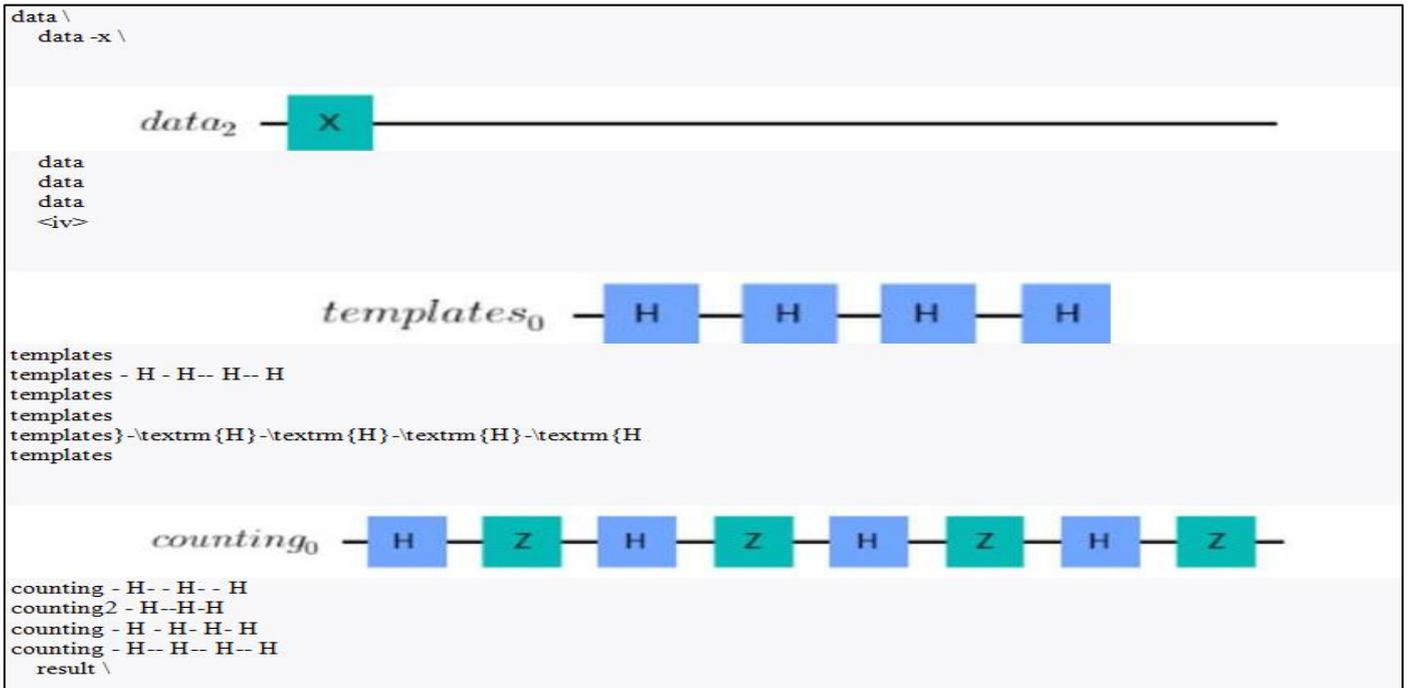

Fig 20 Initializing Registers

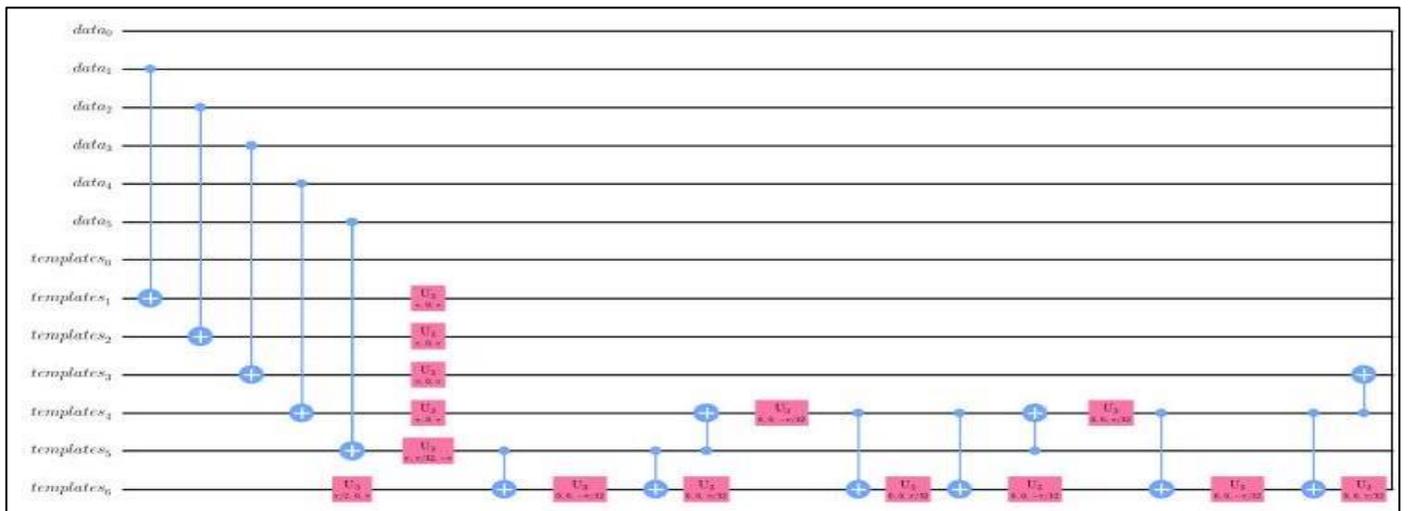

Fig 21 Oracle Section 1

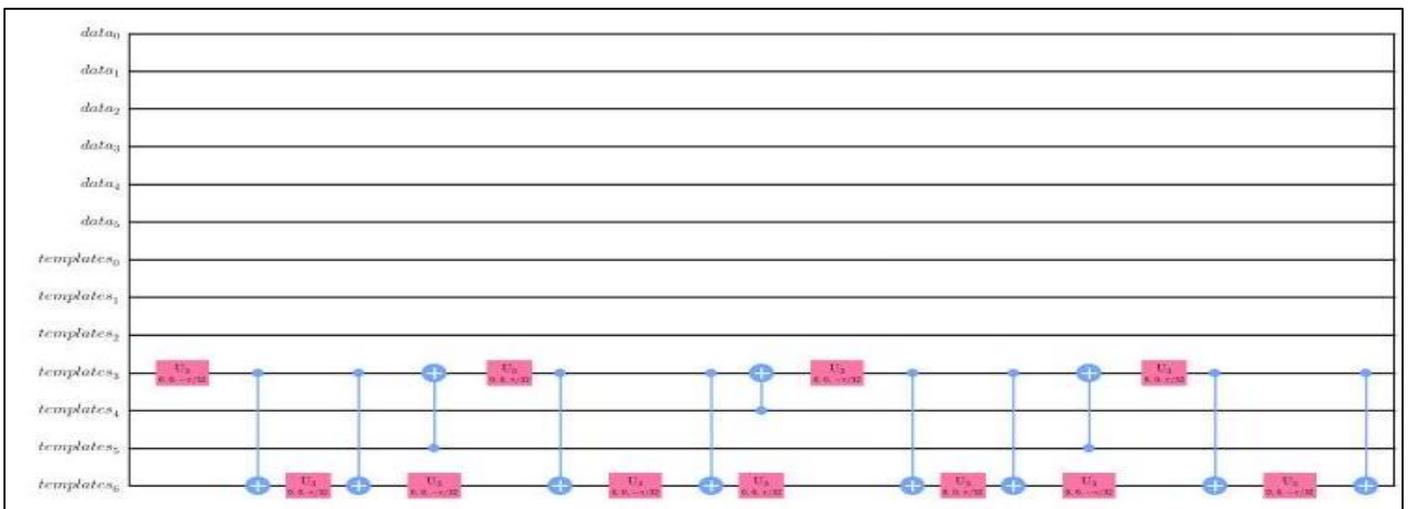

Fig 22 Oracle Section 2

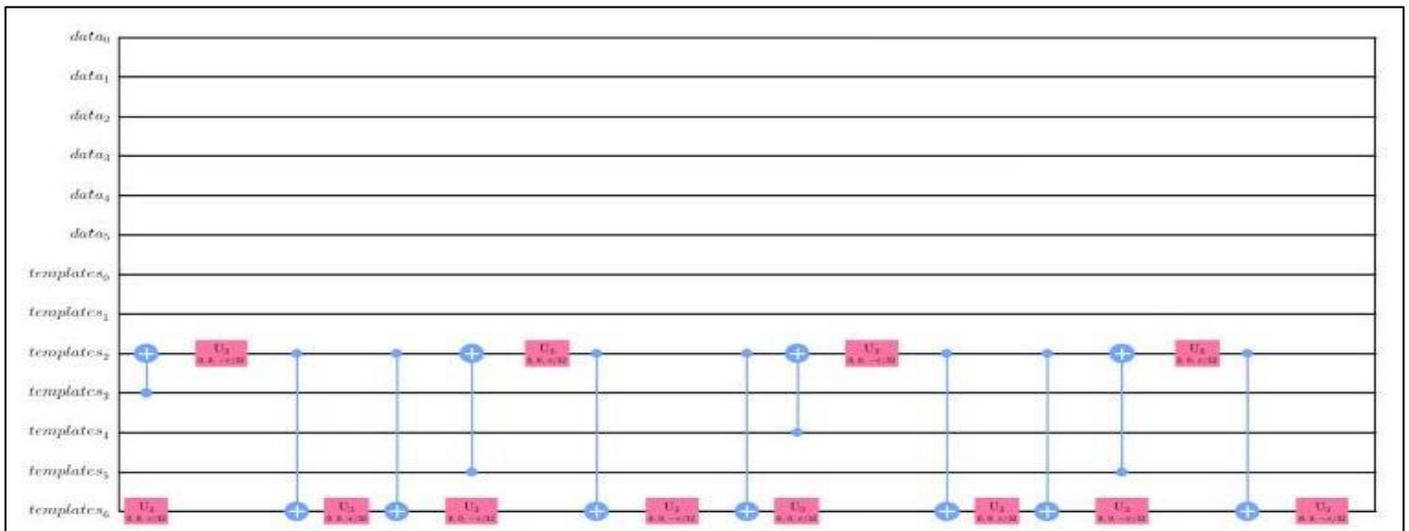

Fig 23 Oracle Section 3

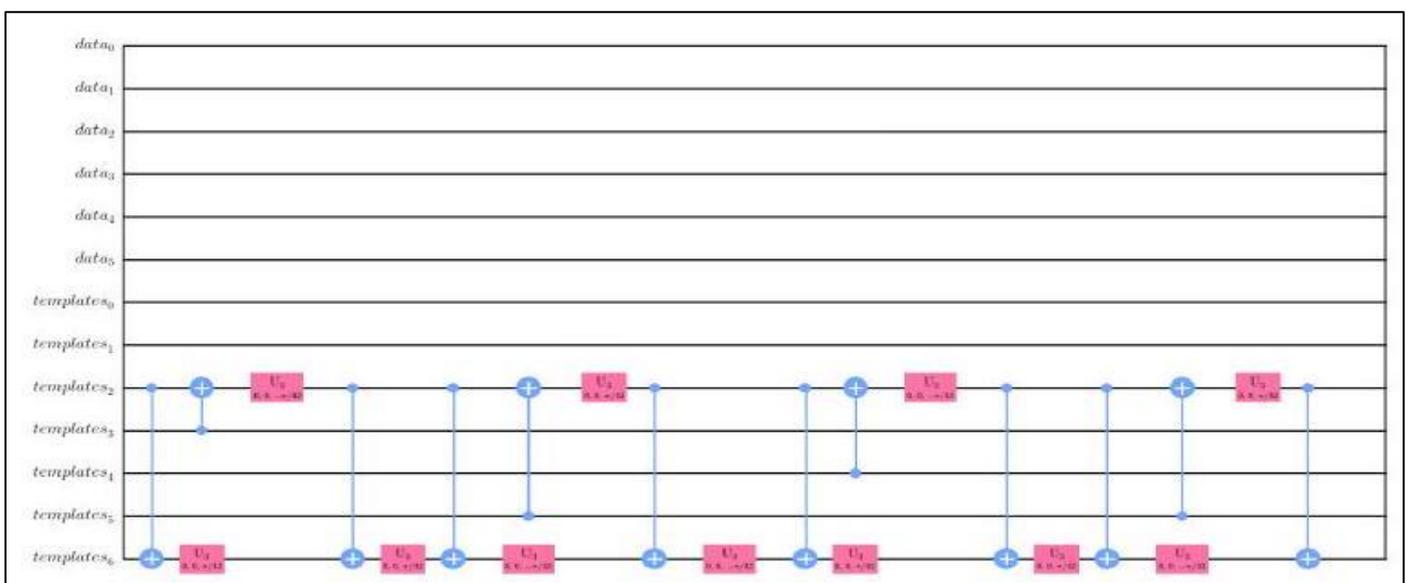

Fig 24 Oracle Section 4

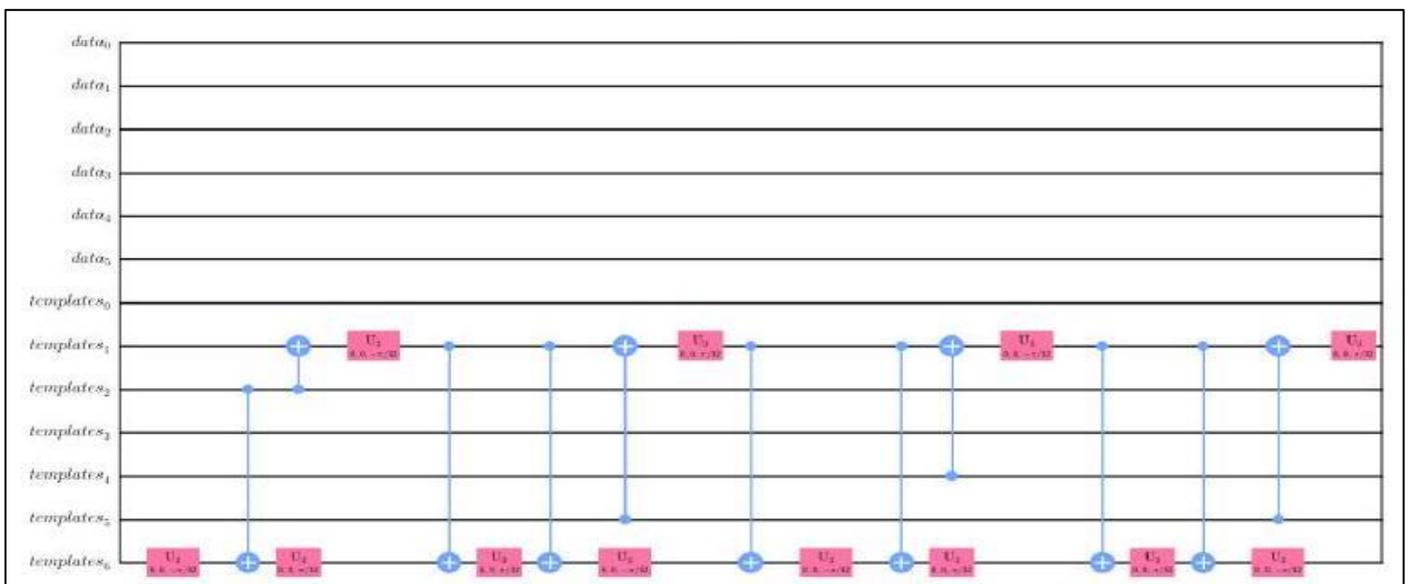

Fig 25 Oracle Section 5

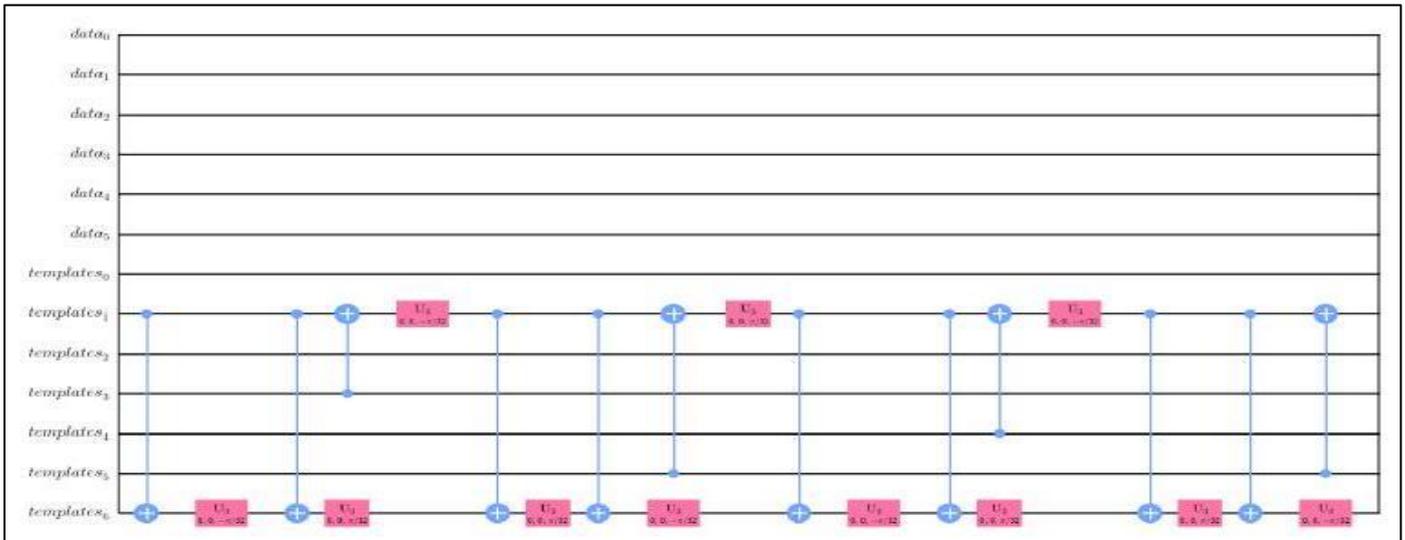

Fig 26 Oracle Section 6

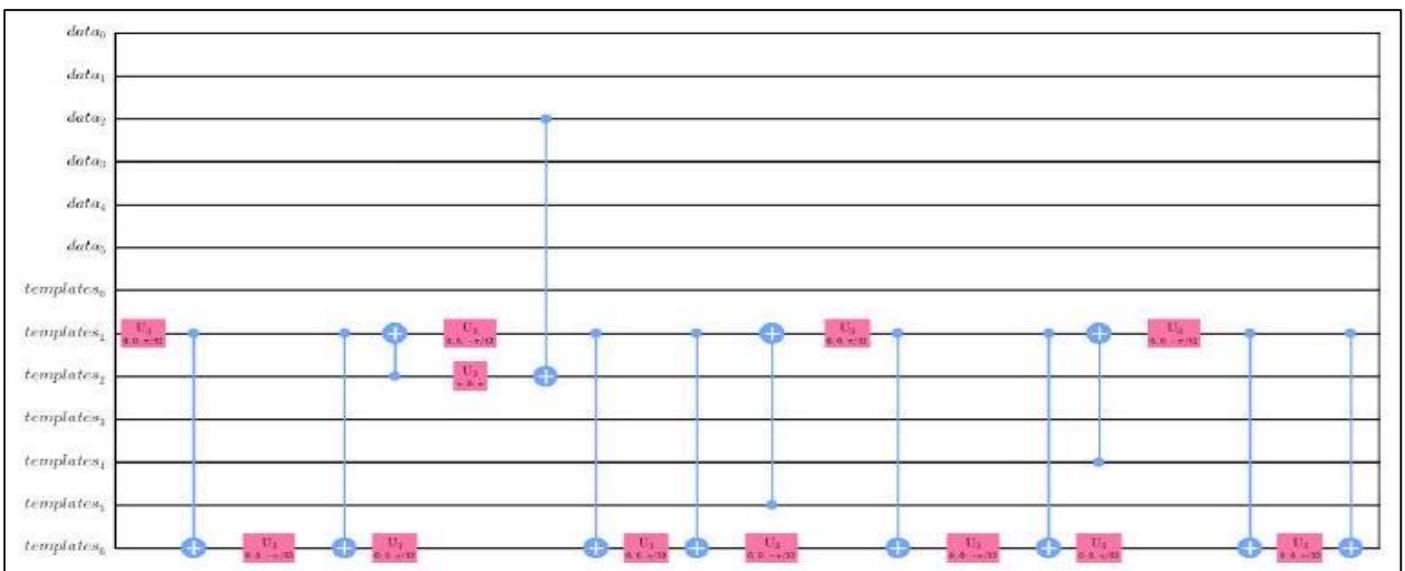

Fig 27 Oracle Section 7

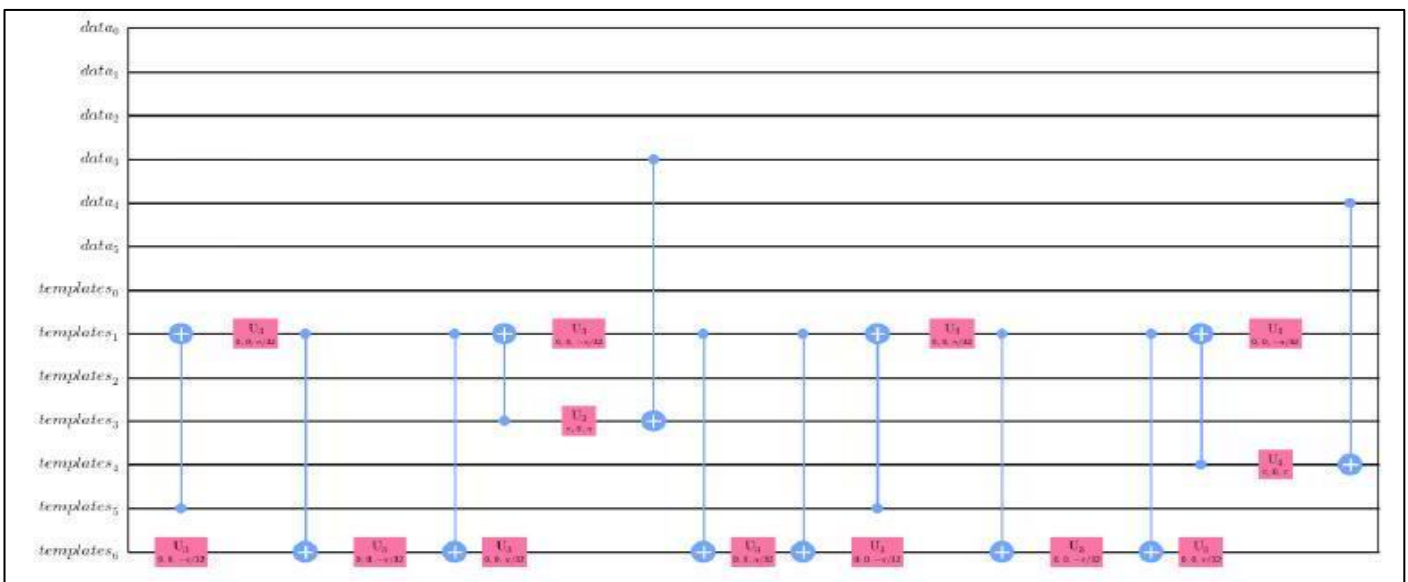

Fig 28 Oracle Section 8

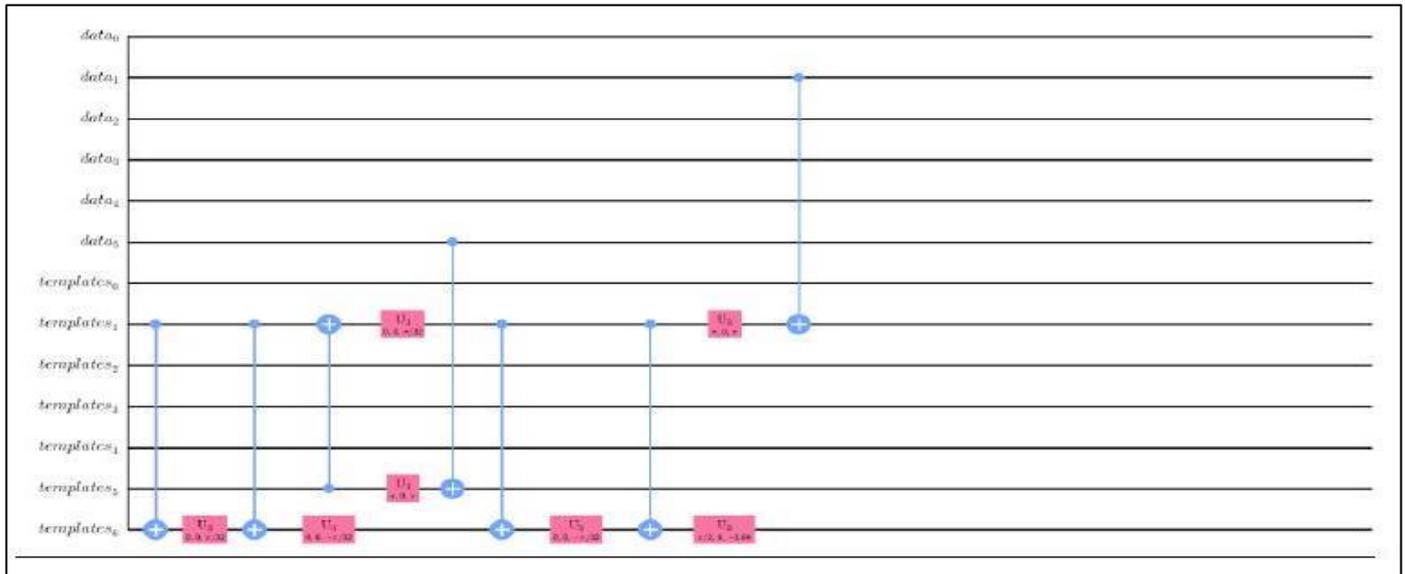

Fig 29 Oracle Section 9

The diffusion operator can be written as:

$$2|\psi\rangle\langle\psi| - \hat{I}_N = \hat{H}^{\otimes n}(2|0\rangle\langle 0| - \hat{I}_N)\hat{H}^{\otimes n}$$

Where $|\psi\rangle$ is the uniform superposition of states and $\hat{I}$ is the N dimensional identity matrix. As $2|\psi\rangle\langle\psi| - \hat{I}_N$ operates a reflection about the $|\psi\rangle$, $2|0\rangle\langle 0| - \hat{I}_N)$ operates a reflection about the $|0\rangle$. It turns out that Grover diffusion can be implemented on a quantum circuit with a phase shift operator that negates all the states except for $|0\rangle$ sandwiched between $\hat{H}^{\otimes n}$ gates.

However, this actually introduce an overall -1 to all states. This would not change the results of Grover's searching part, but will affect the quantum counting part, introducing an overall $\pi$ phase. This is solved by applying a Z-Gate to the lowest qubit in the counting register. (All the other qubits represents powers of two, which will result in the Z-Gate being applyied even times, producing just a 1 rather than -1 ) Third Step involves Quantum Counting where we estimate the number of repetitions needed for multiple matches. It involves generation and application of controlled Grover's gate. We can use .to_gate() and .control() to create a controlled gate from a circuit. We will call our Grover iterator grit and the controlled Grover iterator cgrit. All those controlled gate functions in qiskit is difficult for this situation because they apply to gates, but our grovers operation is only a function. So either we can rewrite this as a gate or we define our controlled funcion. Then we define and apply QFT (Quantum Fourier Transform) and measuring at last.

The measurement of the quantum counting process for six-qubit data matching with a five-qubit counting register is done. The first qubit is ignored to allow for two templates matching.

A measurement of the counting register in the computational basis returns an integer value between 0 and $2^p - 1$, from which we can now extract the desired estimate of the phase. Intuitively, constructive interference occurs for those elements $\{|l'\rangle\}$ for which

$$\frac{\theta}{\pi} - \frac{l'}{2^p} \simeq 0 \text{ or } \frac{\pi - \theta}{\pi} - \frac{l'}{2^p} \simeq 0$$

We will only be interested in cases in which $r \ll N$, and thus $\theta \ll 1$. Therefore, the observed measurement outcome, which we denote $b$, gives an unambiguous estimate of $\theta$, denoted $\theta_*$ as follows:

$$\theta_* = \begin{cases} \dfrac{b\pi}{2^p}, & b \leqslant 2^{p-1} \\ \pi - \dfrac{b\pi}{2^p}, & b > 2^{p-1} \end{cases}$$

The theoretically most probable outcome b in this case, according to Equations above should be either 2 or 30 . The most probable measurement result is 00010 , which in decimal is 2. The results are shown in Fig. 30 and Fig. 31.

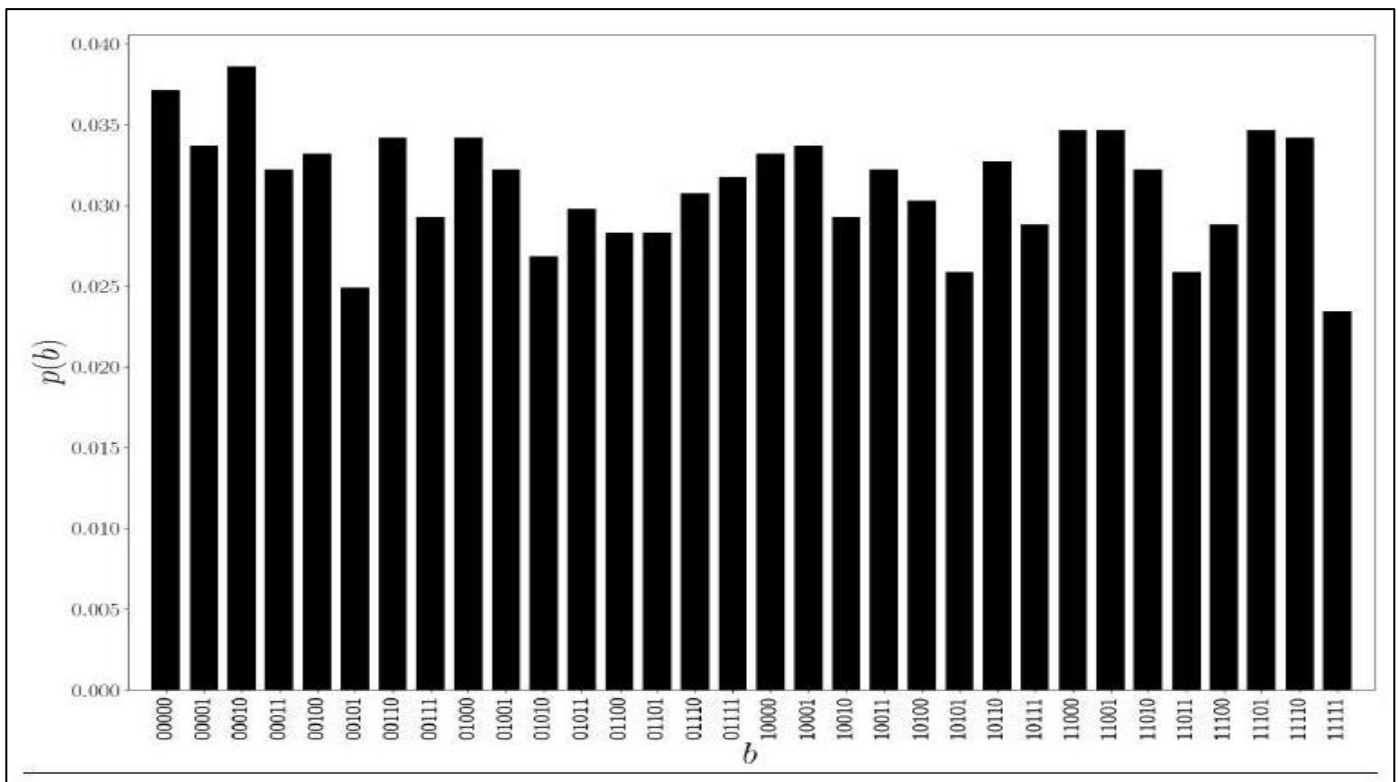

Fig 30 Measurement of Quantum Counting Process

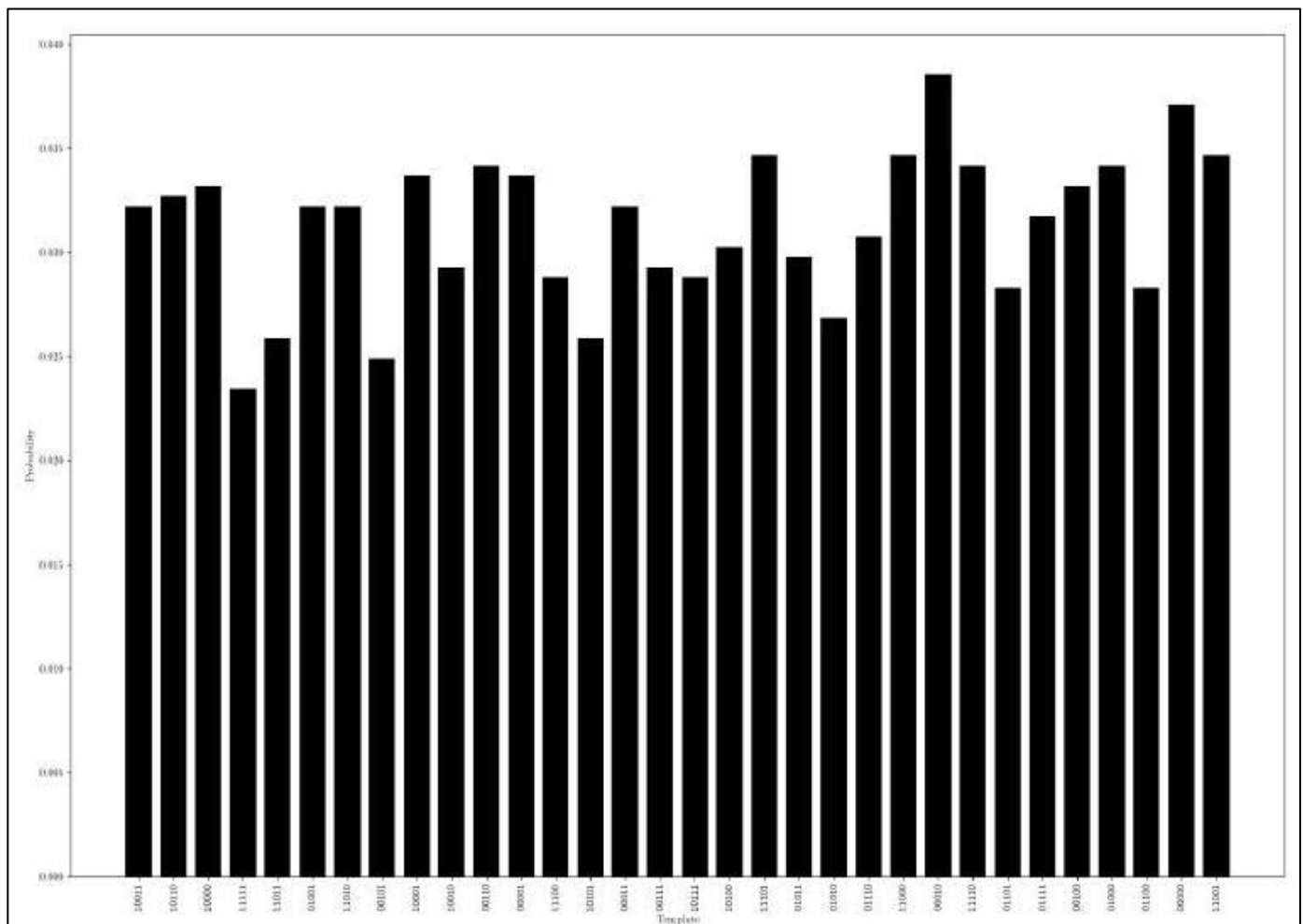

Fig 31 Template Probabilities

Because there are two eigenvalues and we do not know which one is the measure value. We need to run them both and choose the more reasonable one. Then we calculate and search for matched templates.

The measurement of Grover's search process for sixqubit data matching. The data is set as 000111 and the lowest qubit is ignored to allow for two templates matching. With four iterations suggested by the quantum counting process as a numerical output, the two templates that meet the matching criteria are returned with a probability higher than 99% altogether after 2048 trials on the ibmq_qasm_simulator(Fig. 32, Fig. 33).

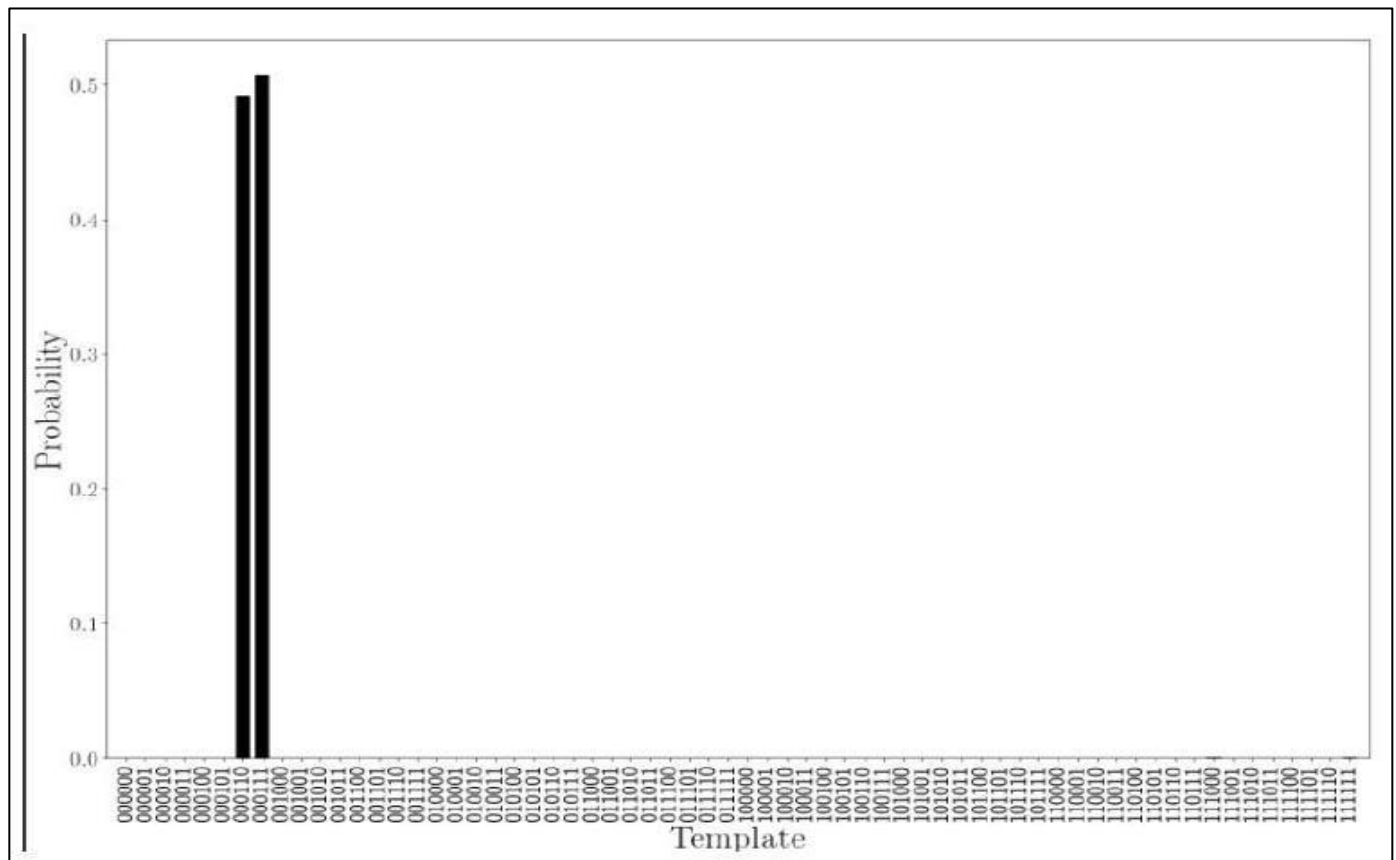

Fig 32 Simulator Probabilities

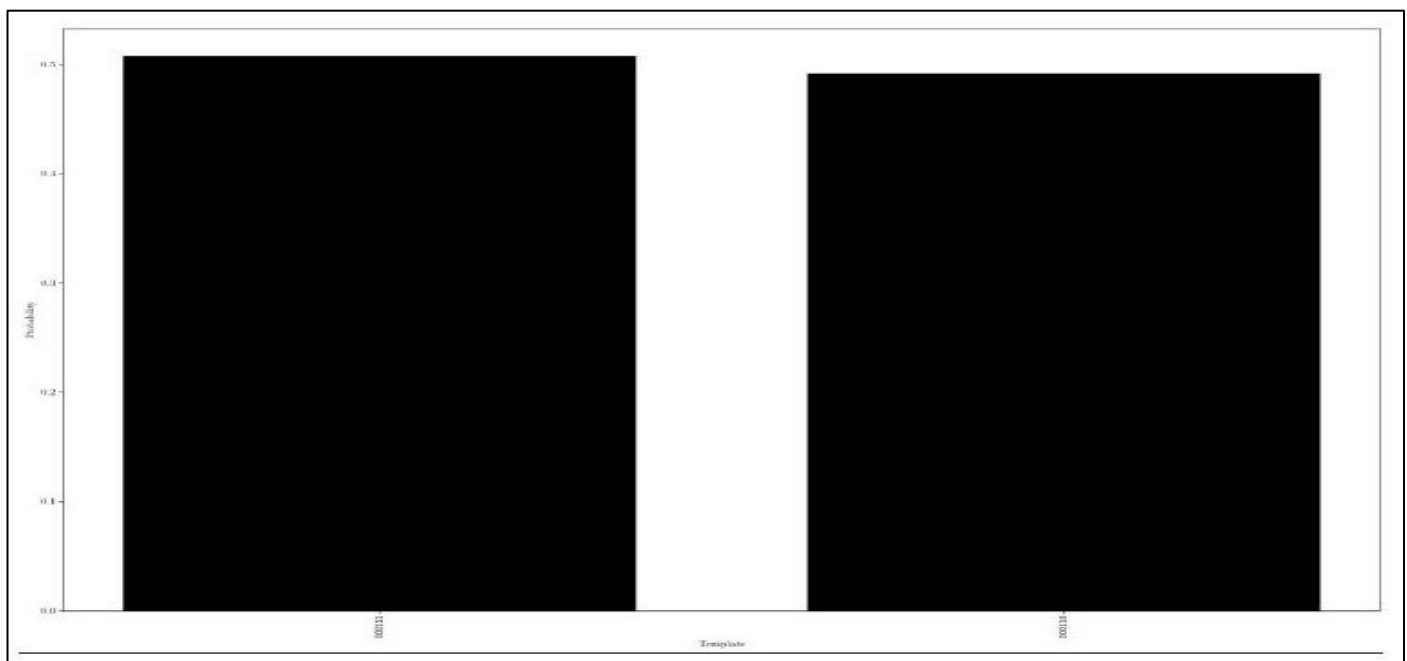

Fig 33 Simulator Probabilities Enlarged

## VI. CONCLUSION AND FUTURE PROSPECTS

The combination of quantum algorithms and quantum metrology presents a promising avenue for advancing the detection of gravitational waves. By utilizing Grover's algorithm for signal detection and quantum metrology for precise parameter estimation, researchers have demonstrated significant improvements in sensor network sensitivity and accuracy. This integrated approach effectively reduces noise, optimizes sensor placement, and enhances network scalability, laying a solid foundation for future gravitational-wave astronomy research. Looking forward, the integration of advanced quantum techniques with gravitational-wave detection holds immense potential for further progress. Continuous refinement of quantum algorithms and metrology techniques is expected to lead to even greater improvements in signal detection efficiency and sensitivity. Future investigations may explore additional quantum algorithms, enhance the scalability of quantum sensor networks, and apply these integrated methodologies to broader areas of observational astronomy. Moreover, advancements in quantum computing hardware will be pivotal in unlocking the full capabilities of these techniques. As these technologies evolve, significant strides are anticipated in expanding our comprehension of the universe and uncovering previously inaccessible cosmic phenomena.

## REFERENCES

[1]. [1] Lloyd, S. (1996). Universal quantum simulators. Science, 273(5278).

[2]. [2] Giovannetti, V., Lloyd, S., Maccone, L. (2004). Quantumenhanced measurements: Beating the standard quantum limit. Science, 306(5700).

[3]. [3] Paris, M. G., Rehacek, J. (2004). Quantum state estimation. Lecture Notes in Physics, 649.

[4]. [4] Childs, A. M., Kothari, R., Somma, R. D. (2018). Quantum algorithm for systems of linear equations with exponentially improved dependence on precision. SIAM Journal on Computing, 47(5).

[5]. [5] Kaubruegger, R., Vasilyev, D. V., Schulte, M., Hammerer, K., Zoller, P. (2021). Quantum variational optimization of Ramsey interferometry and atomic clocks. Physical review X, 11(4), 041045.

[6]. [6] A. Montanaro, Quantum pattern matching fast on average, Algorithmica 77, 16 (2017).

[7]. [7] B. P. Abbott, R. Abbott, T. D. Abbott, S. Abraham, F. Acernese, K. Ackley, C. Adams, R. X. Adhikari, V. B. Adya, C. Affeldt et al., Searches for continuous gravitational waves from 15 supernova remnants and Fomalhaut b with Advanced LIGO, Astrophys. J. 875, 122 (2019).

[8]. [8] M. S. Anis, H. Abraham, R. A. AduOffei, G. Agliardi, M. Aharoni, I. Y. Akhalwaya, G. Aleksandrowicz, T. Alexander, M. Amy, S. Anagolum et al., Qiskit: An open-source framework for quantum computing, Zenodo Version: 0.7.2 (2019), doi:10.5281/zenodo.2562111.

[9]. [9] M. Mosca, Counting by quantum eigenvalue estimation, Theor. Comput. Sci. 264, 139 (2001).

[10]. [10] J. Veitch, V. Raymond, B. Farr, W. Farr, P. Graff, S. Vitale, B. Aylott, K. Blackburn, N. Christensen, M. Coughlin, W. Del Pozzo, F. Feroz, J. Gair, C. J. Haster, V. Kalogera, T. Littenberg, I. Mandel, R. O'Shaughnessy, M. Pitkin, C. Rodriguez, C. Röver, T. Sidery, R. Smith, M. Van Der Sluys, A. Vecchio, W. Vousden, and L. Wade, Parameter estimation for compact binaries with ground-based gravitational-wave observations using the LALInference software library, Phys. Rev. D 91, 042003 (2015).

[11]. [11] G. Ashton, M. Hübner, P. D. Lasky, C. Talbot, K. Ackley, S. Biscoveanu, Q. Chu, A. Divarkala, P. J. Easter, B. Goncharov et al., BILBY: A user-friendly Bayesian inference library for gravitational-wave astronomy, Astrophys. J. Suppl. 241, 27 (2019).

[12]. [12] T. Dal Canton and I. W. Harry, Designing a template bank to observe compact binary coalescences in Advanced LIGO's second observing run, arXiv:1705.01845 (2017).

[13]. [13] S. Dwyer, L. Barsotti, S. S. Y. Chua, M. Evans,

[14]. M. Factourovich, D. Gustafson, T. Isogai, K. Kawabe, A. Khalaidovski, P. K. Lam, M. Landry, N. Mavalvala, D. E. McClelland, G. D. Meadors, C. M. Mow-Lowry, R. Schnabel, R. M. S. Schofield, N. Smith-Lefebvre, M. Stefszky, C. Vorvick, and D. Sigg, "Squeezed quadrature fluctuations in a gravitational wave detector using squeezed light," Opt. Express 21, 19047-19060 (2013).

[15]. [14] Chou, Chen-Kuan Chen, Wei Fwu, Peter Lin, Sung-Jan Lee, Hsuan-Shu Dong, Chen-Yuan. (2008). Polarization ellipticity compensation in polarization second-harmonic generation microscopy without specimen rotation. Journal of biomedical optics. 13. 014005. 10.1117/1.2824379.